\documentclass[prd,nofootinbib,twocolumn]{revtex4}

\pdfoutput=1

\usepackage{color}
\usepackage[squaren]{SIunits}
\usepackage{amsmath}
\usepackage{amsfonts}
\usepackage{slashed}
\usepackage{graphicx}
\usepackage{subfigure}
\usepackage{rotating}

\usepackage{graphicx,amsmath,amssymb,epsfig,verbatim,mathrsfs,array,layout,textcomp,latexsym}
\usepackage{multirow}

\allowdisplaybreaks[1]

\newcommand{\refeq}[1]{Eq.~(\ref{eq:#1})}

\newcommand{\bea}{\begin{eqnarray}}
\newcommand{\eea}{\end{eqnarray}}

\begin{document}

\setlength{\parindent}{0pt}

\hbox{DO-TH 14/17}

\title{$R_K$ and future $b \to s \ell \ell$  BSM opportunities}

\author{Gudrun Hiller}
\affiliation{Institut f\"ur Physik, Technische Universit\"at Dortmund, D-44221 Dortmund, Germany}
\author{Martin Schmaltz}
\affiliation{Physics Department, Boston University, Boston, MA 02215}

\begin{abstract} 
Flavor changing neutral current $|\Delta B|=|\Delta S|=1$ processes are sensitive to possible new physics at the electroweak scale and beyond, providing detailed information about flavor, chirality and Lorentz structure. Recently the LHCb collaboration announced a $2.6 \sigma$ deviation in the measurement of $R_K={\cal{B}}(\bar B \to \bar K \mu \mu)/{\cal{B}}(\bar B \to \bar K ee)$ from the standard model's prediction of lepton universality. We identify dimension six operators which could explain this deviation and study constraints from other measurements. Vector and axial-vector four-fermion operators with flavor structure $\bar s b \bar \ell \ell$ can provide a good 
description of the data. Tensor operators cannot describe the data. Pseudo-scalar and scalar operators only fit the data with some fine-tuning; they can be further probed with the $\bar B \to \bar K ee$ angular distribution.
The data appears to point towards $C_9^{\rm NP \mu } =  -C_{10}^{\rm NP \mu }<0$,  an $SU(2)_{L}$ invariant direction in parameter space supported by $R_K$, the $\bar B \to \bar K^* \mu \mu$ forward-backward asymmetry and the $\bar B_s \to \mu \mu$ branching ratio, which is currently allowed to be smaller than the standard model prediction.
We present two leptoquark models which can explain the FCNC data and give predictions for the LHC and rare decays.
\end{abstract}

\maketitle

\section{Introduction}

At the tree level the Standard Model (SM) has only flavor-universal gauge interactions, all flavor-dependent interactions originate from the Yukawa couplings. The LHCb collaboration recently determined the ratio of  branching ratios of $\bar B \to \bar K \ell \ell$ decays into dimuons over dielectrons \cite{Hiller:2003js},
\begin{align} \label{eq:RK}
R_{K} =\frac{ {\cal{B}}(\bar B \to \bar K \mu \mu) }{{\cal{B}}(\bar B \to \bar K e e) }  \, ,
\end{align}
and obtained
\begin{align} \label{eq:RKdata}
R_{K}^{LHCb} =0.745 \pm^{0.090}_{0.074} \pm 0.036
\end{align}
in the dilepton invariant mass squared bin $1 \, \mbox{GeV}^2 \leq q^2 < 6  \, \mbox{GeV}^2$ \cite{Aaij:2014ora}. Adding statistical and systematic uncertainties in quadrature, this corresponds to a $2.6 \sigma$ deviation from the SM prediction $R_K=1.0003 \pm 0.0001$ \cite{Bobeth:2007dw}, including $\alpha_s$ and subleading $1/m_b$ corrections.
Previous measurements \cite{Wei:2009zv,Lees:2012tva} had significantly larger uncertainties and were consistent with unity. Taken at face value, (\ref{eq:RKdata}) points towards lepton-non-universal physics beyond the Standard Model (BSM).

In this work we discuss model-independent interpretations of the LHCb result for $R_K$, taking into account 
all additional available information on $b \to s \ell \ell$ transitions. We also propose two viable models with leptoquarks which predict $R_K < 1$ and point out which future measurements may be used to distinguish between our models and other possible new physics scenarios.

The plan of the paper is as follows:
In Section \ref{sec:mia} we introduce the low energy Hamiltonian and relevant observables for
$b\to s \ell \ell$ transitions. In Section \ref{sec:ops} we perform a model-independent analysis and
identify higher dimensional operators that can describe existing data. In Section
\ref{eq:particles} we discuss two models in which the flavor-changing neutral current is mediated at tree-level with the favored flavor, chirality and Dirac structure as determined by our model-independent analysis.
We summarize in Section  \ref{sec:con}.

\section{Model-Independent analysis \label{sec:mia}}

To interpret the data we use the following effective $|\Delta B|=|\Delta S|=1$ Hamiltonian
\begin{align}
  \label{eq:Heff}
  {\cal{H}}_{\rm eff}= 
   - \frac{4\, G_F}{\sqrt{2}}  V_{tb}^{} V_{ts}^\ast \,\frac{\alpha_e}{4 \pi}\,
     \sum_i C_i(\mu)  {\cal{O}}_i(\mu) \, ,
\end{align}
where $\alpha_e$, $V_{ij}$ and $G_F$ denote the fine structure constant, the CKM matrix elements and Fermi's constant, respectively. The complete set of dimension six $\bar s b \ell \ell$ operators comprises V, A operators (referring to the lepton current)
\begin{align} 
  {\cal{O}}_{9} & =  [\bar{s} \gamma_\mu P_{L} b] \, [\bar{\ell} \gamma^\mu \ell] \,, \quad
  {\cal{O}}_{10}  = [\bar{s} \gamma_\mu P_{L} b] \, [\bar{\ell} \gamma^\mu \gamma_5 \ell] \,, 
\end{align}
S, P operators
\begin{align} 
  {\cal{O}}_{S} & =  [\bar{s}  P_{R} b] \, [\bar{\ell} \ell] \,, \quad
  {\cal{O}}_{P}  = [\bar{s}  P_{R} b] \, [\bar{\ell}  \gamma_5 \ell ]\,, 
\end{align}
and tensors
\begin{align}
   {\cal{O}}_T   & = [\bar{s} \sigma_{\mu\nu} b] \, [\bar{\ell} \sigma^{\mu\nu} \ell]\,, \quad
   {\cal{O}}_{T5}  = 
    [\bar{s} \sigma_{\mu\nu} b]\, [\bar{\ell} \sigma^{\mu\nu}\gamma_5 \ell] \,.
\end{align}

Chirality-flipped operators $ {\cal{O}}^\prime$ are obtained by inter-changing the chiral projectors $P_L \leftrightarrow P_R$ in the quark currents.

Parity conservation of the strong interactions implies that $\bar B_s \to \ell \ell$ decays depend on the Wilson coefficient combinations $C_- \equiv C -C^\prime$,
whereas $\bar B \to \bar K \ell \ell$ decays depend on $C_+ \equiv C +C^\prime$. There are no tensor or vector $C_9^{(\prime)}$ contributions to $\bar B_s \to \ell \ell$ decays.

The SM predicts $C_{9}=-C_{10}=4.2$ at the $m_b$-scale, universally for all leptons. All other semileptonic Wilson coefficients are negligible. We can use this fact to simplify our notation: in the following $C_9^{\rm SM}$ and $C_{10}^{\rm SM}$ denote the SM contributions to $C_9$ and $C_{10}$ whereas $C_9^{\rm NP}$ and $C_{10}^{\rm NP}$ denote possible new physics contributions. For all other Wilson coefficients we omit the ${}^{\rm NP}$ superscript because non-negligible contributions are necessarily from new physics.
To discuss lepton-non-universality, we add a lepton flavor index to the operators and their Wilson coefficients.\footnote{We do not consider lepton flavor violation in this paper.}

We continue by listing the most relevant measurements which provide constraints on the Wilson coefficients.
All errors are 1 $\sigma$ unless stated otherwise.
The average time-integrated branching fraction of $\bar B_s \to \ell \ell$ decays,
with recent data \cite{bsmmcombi,Beringer:1900zz} and SM predictions  \cite{Bobeth:2013uxa} is
\begin{align}
{\cal{B}}(\bar B_s \to  ee)^{\rm exp} & <  2.8 \cdot 10^{-7}  \, , \\
{\cal{B}}(\bar B_s \to  \mu \mu )^{\rm exp} & = (2.9 \pm 0.7 )\cdot 10^{-9} \, ,\\
{\cal{B}}(\bar B_s \to  ee)^{\rm SM} & =  (8.54  \pm 0.55) \cdot 10^{-14} \, ,\\
{\cal{B}}(\bar B_s \to  \mu \mu )^{\rm SM} & = ( 3.65 \pm 0.23) \cdot 10^{-9}
\end{align}
resulting in 
\begin{align}  \label{eq:Bsee}
\frac{ {\cal{B}}(\bar B_s \to  ee)^{\rm exp} }{{\cal{B}}(\bar B_s \to ee)^{\rm SM} }& < 3.3 \cdot 10^6\, , \\
 \label{eq:Bsmm}
\frac{ {\cal{B}}(\bar B_s \to  \mu \mu)^{\rm exp} }{{\cal{B}}(\bar B_s \to \mu \mu)^{\rm SM} } & =
0.79 \pm 0.20 \, .
\end{align}
Ratios (\ref{eq:Bsee}), (\ref{eq:Bsmm}) yield model-independent constraints on
\begin{align} \nonumber 
\frac{ {\cal{B}}(\bar B_s \!\!\to\!  \ell \ell) }{{\cal{B}}(\bar B_s\! \!\to\! \ell \ell)^{\rm SM} } &\! =\!
   |1\!\!-\!0.24 (C_{10}^{\ell \rm NP} \!\!-\!C_{10}^{ \ell \prime})\!-\!y_\ell C_{P-}^{\ell} |^2
    \!+\!|y_\ell C_{S-}^{\ell }  |^2  \label{eq:Bsmm2}  \\
y_\mu= \,& 7.7 , \ ~y_e=(m_\mu/m_e) y_\mu=1.6 \cdot 10^3\, . 
\end{align}

We further employ the $\bar B \to \bar K ee $ branching ratio recently measured by LHCb \cite{Aaij:2014ora}. This is currently the most precise determination and uses data with $1 \, \mbox{GeV}^2 \leq q^2 < 6  \, \mbox{GeV}^2$
\begin{align} \nonumber 
{\cal{B}}(\bar B \to \bar K ee)^{ LHCb} & = (1.56^{+0.19 + 0.06}_{-0.15-0.04}) \cdot 10^{-7} \, ,\\
{\cal{B}}(\bar B  \to \bar K ee)^{\rm SM} & = (1.75^{+0.60}_{-0.29})  \cdot 10^{-7}\, , \\
\label{eq:BKee}
\frac{{\cal{B}}(\bar B \to \bar K ee)^{ LHCb}}{{\cal{B}}(\bar B  \to \bar K ee)^{\rm SM}} & = 0.83 \pm 0.21\, .
\end{align}
Here the SM prediction is taken from \cite{Bobeth:2012vn} and in the ratio we added uncertainties in quadrature and symmetrized.

We also use the branching ratios of  inclusive $\bar B \to X_s \ell \ell $ decays
for $q^2 > 0.04 \, \mbox{GeV}^2$ \cite{hfag}
\begin{align} \nonumber 
{\cal{B}}(\bar B \to X_s  ee)^{\rm exp} & = (4.7 \pm 1.3)  \cdot 10^{-6}  \, ,\\ 
\label{eq:BXsll}
{\cal{B}}(\bar B  \to X_s  \mu \mu )^{\rm exp} & = (4.3 \pm 1.2 ) \cdot 10^{-6}\, , \\
{\cal{B}}(\bar B  \to X_s  \ell \ell  )^{\rm SM} & = (4.15 \pm 0.70 ) \cdot 10^{-6}\, , \quad \ell=e,\mu \, , \nonumber 
\end{align}
where the SM prediction is taken from \cite{Ali:1999mm}.

The observables $F_H^\ell$ and $A_{FB}^\ell$ in the $\bar B \to \bar K \ell \ell$ angular distribution 
\begin{align} \label{eq:angular}
\frac{1}{\Gamma^\ell}\! \frac{d\Gamma^\ell}{d\! \cos\! \theta_\ell} \!=
      \!\frac{3}{4}(1\!-\!F_H^\ell\!) (1\!-\!\cos^2\! \theta_\ell\!) \!
      +\!\frac{F_H^\ell}{2}\!+\!A_{\rm FB}^\ell \cos \theta_\ell 
\end{align}
are sensitive to S, P and T operators and related to $R_K$  \cite{Bobeth:2007dw}. 
Here, $\Gamma^\ell$ denotes the decay rate and $\theta_\ell$ the angle between the
negatively charged lepton with respect to the $\bar B$ in the dilepton center of mass system. When no S, P or tensors are present%
\footnote{More precisely, contributions to $F_H^\ell$ from (axial) vectors are proportional $ m_\ell^2/q^2$ \cite{Bobeth:2007dw} and too small to be observable given projected uncertainties for $\ell=e,\mu$.}
the angular distribution is SM-like with $F_H^\ell, A_{\rm FB}^\ell =0$.
Current data on $F_H^\mu$ and $A_{\rm FB}^\mu$ are consistent with the SM \cite{Aaij:2012vr,Aaij:2014tfa}
and provide useful BSM constraints \cite{Bobeth:2012vn} which we will use in Section \ref{sec:opsC}.
The electron angular observables $F_H^e$ and $A_{\rm FB}^e$ have not been measured yet but they will eventually be important for distinguishing between different possible BSM explanations of $R_K$.

\section{Interpretations with operators \label{sec:ops}}

We explore which of the four-fermion operators in \refeq{Heff} can accommodate the data on $R_K$ (\ref{eq:RKdata}) as well as all the other $b \to s \ell \ell$, $\ell=e,\mu$ constraints.
We study  (axial)-vectors, (pseudo-)scalars and tensors in Sections \ref{sec:opsA}, \ref{sec:opsB} and
 \ref{sec:opsC}, respectively and summarize in  \ref{sec:opsD}.

\subsection{ (Axial)-vectors  \label{sec:opsA}}

{}Following  \cite{Das:2014sra}, the $R_K$ data implies at 1 sigma  
\begin{align} \label{eq:Xlimit}
& 0.7 \lesssim {\rm Re}[X^e-X^\mu] \lesssim  1.5 \, , \\
X^\ell&=C_9^{{\rm NP} \ell} +C_9^{\prime  \ell}-
(C_{10}^{{\rm NP}  \ell} +C_{10}^{\prime \ell}) \,  , \quad \ell=e, \mu \, .
\end{align}

Global fits to radiative, leptonic, and semileptonic $b\to s$ transitions which includes the wealth of recent $\bar B \to \bar K^* (\to \bar K \pi) \ell \ell$ data have been performed  by several groups \cite{Descotes-Genon:2013wba,Altmannshofer:2013foa,Beaujean:2013soa} assuming contributions from V,A and primed operators only.
The fits assume lepton-universality but the dominant data are from hadron colliders and
hence the results apply to the muonic $C_i^{(\prime)\mu}$ coefficients to a very good approximation.

We discuss generic features of the fits.
Axial vector operators: All groups find that only small BSM contributions are allowed ${\rm Re}[C_{10}^{{\rm NP} \mu}, C_{10}^{\prime \mu}] \sim [-0.4 \ ... +0.1]$. This is too small to explain $R_K$ 
without additional contributions from other operators or from electron modes, see Eq.~(\ref{eq:Xlimit}). Moreover, the contributions with the largest allowed magnitude, $C_{10}^{(\prime) \mu}\sim -0.4$, have the wrong sign to help in Eq.~(\ref{eq:Xlimit}). 
Vector operators: Global fits which also include $\bar B_s \to \phi \mu \mu$ data \cite{Horgan:2013pva} indicate sizable contributions from vector operators $O_9^{(\prime) \mu}$. In fact, $C_9^{{\rm NP} \mu} \sim -1$ is found to have the right sign and magnitude to explain $R_K$. However, most fits find that $C_9^{\prime \mu}$
is of similar size and opposite in sign so that the contributions to
$R_K$ in \refeq{Xlimit} cancel. Again, other operators or electrons are needed.
To summarize, at this point the outcome of the global fits (performed without taking into account $R_K$) is inconclusive, whether or not BSM physics is preferred by the data depends on how hadronic uncertainties are treated and on the data set chosen. While the SM gives a good fit \cite{Beaujean:2013soa} all groups indicate an intriguing support for sizable  $C_{9}^{(\prime) \rm NP}$, triggered by LHCb's paper \cite{Aaij:2013qta}. Future updates including the analysis of the $3 \rm {fb}^{-1}$ data set will shed light on this.

For our UV-interpretation of the data in the next Section \ref{eq:particles}
it is useful to change from the  $\mathcal{O}_{9,10}^{(\prime) \ell}$ basis to one with left- and right projected leptons
\begin{align}
\mathcal{O}_{LL}^\ell&\equiv (\mathcal{O}_9^\ell-\mathcal{O}_{10}^\ell)/2 \, , \quad \mathcal{O}_{LR}^\ell \equiv (\mathcal{O}_9^\ell+\mathcal{O}_{10}^\ell)/2 \, , \\  \mathcal{O}_{RL}^\ell& \equiv (\mathcal{O}_9^{\prime \ell}-\mathcal{O}_{10}^{\prime \ell})/2 \, ,  \quad \mathcal{O}_{RR}^\ell \equiv (\mathcal{O}_9^{\prime \ell}+\mathcal{O}_{10}^{\prime \ell})/2 \, ,
\end{align}
therefore 
\begin{align}
C_{LL}^\ell & = C_9^\ell-C_{10}^\ell \, , \quad C_{LR}^\ell = C_9^\ell+C_{10}^\ell \, , \\
C_{RL}^\ell & = C_9^{\prime \ell}-C_{10}^{\prime \ell} \, , \quad C_{RR}^\ell = C_9^{\prime \ell}+C_{10}^{\prime \ell} \, .
\end{align}
If we assume new physics in muons alone we can rewrite
Eqs.~(\ref{eq:Bsmm2}) and (\ref{eq:Xlimit}) to obtain constraints on the BSM contributions
\bea
0.0& \lesssim & {\rm Re}[C_{LR}^{\mu}+C_{RL}^{\mu}-C_{LL}^{\mu}-C_{RR}^{\mu}] \lesssim 1.9  \, , \nonumber \\
0.7& \lesssim & -{\rm  Re}[C_{LL}^{\mu}+C_{RL}^{\mu}]  \lesssim 1.5 \ .
\eea
One sees that the only single operator which improves both constraints is $\mathcal{O}_{LL}^\mu$ and a good fit  of the above is obtained with 
\begin{align} \label{eq:benchmark}
C_{LL}^\mu \simeq -1 \, , 
\quad  C_{ij}^\mu=0 ~ {\rm otherwise} \, 
\end{align}
which we adopt as our benchmark point.
In terms of the standard basis, this choice implies $C_9^{\rm NP \mu}=-C_{10}^{\rm NP \mu}\simeq -0.5 $ and  $C_9^{\rm NP \mu}+ C_{10}^{\rm NP \mu}=0$.
It would be interesting to perform global fits as in \cite{Descotes-Genon:2013wba,Altmannshofer:2013foa,Beaujean:2013soa} with this constraint to probe
how this scenario stacks up against all $|\Delta B|=|\Delta S|=1 $ data. In particular, all transversity amplitudes corresponding to $\bar \ell \gamma^\mu (1+\gamma_5) \ell$ currents $(A^R)$  in $\bar B \to \bar K^* (\to \bar K \pi) \ell \ell$ decays  in this
scenario remain SM-valued.

A few comments are in order:
If $\bar B_s \to \mu \mu$ data had shown an enhancement (of similar size for concreteness) over the SM, the preferred one-operator benchmark would have been $C_{RL}^\mu \simeq -1$ with all other coefficients vanishing. In that case the new physics would have to generate right-handed quark FCNCs instead of SM-like 
left-handed ones. This fact that $\bar B_s \to \mu \mu$ is a diagnostic for the chirality of the quarks in BSM FCNCs makes more precise measurements of $\bar B_s \to \mu \mu$ especially interesting.
Second, the constraint $C_9^{\rm NP \mu } +  C_{10}^{\rm NP \mu }=0$ which is motivated by $SU(2)_L$-invariance of the UV physics ensures that the combination 
${\rm Re} [C_9 C_{10}^*]/(|C_9|^2+|C_{10}|^2)$ remains invariant, {\it i.e.} SM-valued. This is helpful because this combination enters in the dominant contributions to the forward-backward asymmetry as well as in the angular observable $P_5^\prime$ in $\bar B \to \bar K^* \mu \mu$ decays at high-$q^2$, where data are in agreement with the SM \cite{Hambrock:2013zya}. In fact, all high $q^2$ observables driven by $\rho_2/\rho_1$ follow this pattern of Wilson coefficients \cite{Bobeth:2012vn} and would remain invariant if $C_{LL}^\mu \neq 0$ were the sole BSM effect.
Third,  $C_{LL}^\mu <0$ shifts the location of the zero which is present in $A_{\rm FB}(\bar B \to \bar K^* \mu \mu) $ at low $q^2$ to higher values, also in agreement with current data.

\subsection{(Pseudo)scalars  \label{sec:opsB}}

Following  \cite{Bobeth:2007dw} the $R_K$-data implies for (pseudo-) scalar contributions at 1 sigma \footnote{\label{foot-label}In the evaluation of the S, P and T, T5 constraints we keep corrections proportional to a single power of the muon mass.}
\begin{align}
15\lesssim 2 {\rm Re}[C_{P+}^\mu ]\!-\! |C_{S+}^{\mu}|^2 \!-\! |C_{P+}^{\mu}|^2 
\!+\! |C_{S+}^{e}|^2\! +\!| C_{P+}^{e}|^2 \! \lesssim\! 34 \, .
\end{align}
This constraint cannot be satisfied with muon operators because the coefficients of the quadratic terms enter with minus signs and the linear term is either too small or dominated by the quadratic terms.
In addition, muon scalars are subject to the $\bar B_s \to \mu \mu$ constraint (\ref{eq:Bsmm}), (\ref{eq:Bsmm2}) 
\begin{align} 
  \label{eq;Sboundmu}
   | C_{P-}^\mu |& \lesssim 0.3 \, , |C_{S-}^\mu| \lesssim 0.1 \quad {\cal{B}}(\bar B_s \to \mu \mu) \,   .
\end{align}

The corresponding electron contributions are bounded by (\ref{eq:BKee}).
We obtain at $1 \sigma (2 \sigma)$
\begin{align} 
  \label{eq;Sbounde}
   |C_{S+}^e|^2 + |C_{P+}^e|^2 & \lesssim 4 \, (24)  \quad {\cal{B}}(\bar B \to \bar K ee) \, .
\end{align}
The constraints from inclusive decays (\ref{eq:BXsll}) are weaker, and 
do not  involve interference terms
\begin{align} 
  \label{eq;Sbounde2}
  |C_{S}^e|^2 + |C_{P}^e|^2+
  |C_{S}^{\prime e}|^2 + |C_{P}^{\prime e}|^2 & \lesssim 53  \, (91)  \quad {\cal{B}}(\bar B \to X_s ee)  \, .
\end{align}
We checked that the available data on inclusive decays in the bin $1 \, \mbox{GeV}^2 < q^2 < 6 \, \mbox{GeV}^2$ is even less constraining.

We learn that at 1 $\sigma$ an explanation of $R_K$ by (pseudo-) scalar operators is excluded. At 2 $\sigma$ this is an option if the electron contributions are sizable.
However, in this case one needs to accept cancellations between $C_{S,P}^{e}$ and $C_{S,P}^{\prime e}$ due to the $\bar B_s \to ee$ constraint
 (\ref{eq:Bsee}), (\ref{eq:Bsmm2}) 
\begin{align} 
  \label{eq;Bseebound}
   |C_{S-}^e|^2 + |C_{P-}^e|^2 & \lesssim 1.3  \quad {\cal{B}}(\bar B_s \to  ee) \, .
\end{align}
In any case, a measurement of the flat term $F_H^e$ in the $\bar B \to \bar K ee$ angular distribution (\ref{eq:angular}) would probe this scenario. This fact, that $R_K$ and $F_H^e$ are correlated had already been pointed out in \cite{Bobeth:2007dw}.

\subsection{Tensors \label{sec:opsC}}

Following  \cite{Bobeth:2007dw} the $R_K$-data
implies for tensor contributions at 1 sigma \textsuperscript{\ref{foot-label}}
\begin{align}
5 \lesssim-2 {\rm Re}[C_T^\mu]-  |C_T^{\mu}|^2 - |C_{T5}^{\mu}|^2 + |C_T^{e }|^2 +| C_{T5}^{e }|^2   
\lesssim 11 \, .
\end{align}

We see that the contributions from muon tensors have the wrong sign to help satisfy the inequalities. Moreover, their magnitudes are strongly constrained  by
the measurement of the flat term in the $\bar B \to \bar K \mu \mu$ angular distributions, $F_H^\mu$, see Eq.~(\ref{eq:angular})  at LHCb
at $95\,\%$ CL \cite{Bobeth:2012vn}
\begin{align} 
  \label{eq;Tboundmu}
  |C_T^\mu|^2 + |C_{T5}^\mu|^2 & \lesssim 0.5 \, .
\end{align}
Tensor contributions in the electron modes are currently best constrained by inclusive decays
and by (\ref{eq:BKee}).
We obtain at $1 \sigma (2 \sigma)$
\begin{align} 
  \label{eq;Tbounde}
  |C_T^e|^2 + |C_{T5}^e|^2 & \lesssim 1.1  \, (1.9)  \quad {\cal{B}}(\bar B \to X_s ee)  \, ,\\
   |C_T^e|^2 + |C_{T5}^e|^2 & \lesssim 1.3  \, (8)  \quad {\cal{B}}(\bar B \to \bar K ee) \, .
\end{align}

We conclude that the current data on $R_K$ cannot be explained with new physics in tensor operators alone.

\subsection{Summary of model-independent constraints  \label{sec:opsD}}

Excluding solutions which require more than one type of operator from our list S,P,A,V,T, flipped ones, both lepton species, we obtained three possible $R_K$ explanations:

\begin{itemize}
\item[\it i)] V,A muons
\item[\it ii)] V,A electrons
\item[\it iii)]  S,P electrons (disfavored at 1 $\sigma$ and requires cancellations, testable with $\bar B \to \bar K ee$ angular distributions)
\end{itemize}

In the next section we present example models with (multi)-TeV mass particles which realize the V,A scenarios.

\section{Two simple leptoquark models \label{eq:particles}}

{}Flavor violating current-current operators can be generated at the tree level by integrating out new particles with flavor violating couplings. One possibility is a neutral spin-1 particle with flavor-changing quark couplings and non-universal couplings to muons and electrons \cite{Altmannshofer:2014cfa}. 

Here we pursue a different avenue and consider a scalar leptoquark $\phi$. By choosing specific flavor-violating couplings for the leptoquark one can arrange for it to generate (axial) vector current-current operators which can explain $R_K$.

We present two models, one with new physics coupling to electrons (Section \ref{sec:RLe}) and one with new physics coupling to muons (Section \ref{sec:LLmu}) .

\subsection{A model with a $RL$ operator for electrons \label{sec:RLe}}

For example, consider $\phi$ to have mass $M$ and transform as $(3,2)_{1/6}$ under $(SU(3), SU(2))_{U(1)}$ with couplings of the form 
\bea
\label{eq:eyukawa}
\mathcal{L}=-\lambda_{d\ell}\, \phi\, (\bar{d} P_L \ell) \ ,
\eea
where $d$ stands for an unspecified down-type quark (we will choose both $b$ and $s$) and $\ell$ is a lepton doublet.\footnote{Note that the quantum numbers of $\phi$ allow it to be the scalar superpartner of a left-handed quark doublet. Thus the coupling in \refeq{eyukawa} exactly corresponds to one of the R-parity violating couplings that can be added to the superpotential of the MSSM. If, for example, $\phi$ is a third generation squark doublet, then the couplings $\lambda_{de}$ in (\ref{eq:eyukawa}) correspond to $\lambda'_{1d3}$ in the standard R-parity violation notation where $d=2,3$ for the $s$ and $b$ quark. The mass $M$ of a 3rd generation squark might be expected to be near the weak scale in natural supersymmetry or a loop factor above as in split supersymmetry.}
Integrating out $\phi$ at the tree level generates the operator 
\bea
{\cal{H}}_{\rm eff} &=& 
-\frac{|\lambda_{d\ell}|^2}{M^2}  (\bar{d} P_L \ell)\, (\bar{\ell} P_R d) \nonumber \\
                            &= &
 \frac{|\lambda_{d\ell}|^2}{2M^2}  [\bar{d} \gamma^\mu P_{R} d] \, [\bar{\ell}  \gamma_\mu P_L \ell] \, ,
\eea
where the equality follows from Fierz rearrangement.
By choosing a particular flavor structure in \refeq{eyukawa} we can turn on the Wilson coefficient for the operator which we desire. We choose two non-zero couplings 
\bea
\mathcal{L}=-\lambda_{be}\, \phi\, (\bar{b} P_L \ell_e) -\lambda_{se}\, \phi\, (\bar{s} P_L \ell_e) \, ,
\label{eq:wunderbar}
\eea
to obtain quark-flavor preserving operators and $C_{RL}^e$
\bea
\label{eq:yukawabs}
{\cal{H}}_{\rm eff}= 
 \frac{\lambda_{se}\lambda_{be}^*}{2M^2}  [\bar{s} \gamma^\mu P_{R} b] \, [\bar{\ell_e}  \gamma_\mu P_L \ell_e] \, .
\eea
Comparing to the standard operator basis \refeq{Heff} this gives
\bea
C_{10}'^{e}=-C_9'^e&=&
    \frac{\lambda_{se}\lambda_{be}^*}{V_{tb}V_{ts}^*}\frac{\pi}{\alpha_{e}}\frac{\sqrt{2}}{4 M^2 G_F} \nonumber \\
& =&-\frac{\lambda_{se}\lambda_{be}^*}{2M^2} (24 {\rm TeV})^2
\eea
for electrons. We see that we can fit the experimental value for $R_K$, \refeq{Xlimit}, with
$C_9'^e=-C_{10}'^e\simeq 1/2$ or $M^2/\lambda_{se}\lambda_{be}^*\simeq(24 {\rm TeV})^2$.

We now determine the range of leptoquark masses and couplings which are allowed by other experimental constraints to see if our model is viable.

First off, leptoquarks can be produced in pairs at the LHC from the strong interactions and if they are within kinematic reach, they yield easily identifiable $\ell\ell j j$ signatures. Current lower bounds on leptoquark masses depend on the  flavor of the leptoquark and range from 500 GeV to 1 TeV \cite{Beringer:1900zz,CMS:2014qpa,ATLAS:2013oea}. To be conservative in establishing the viability of our scenario we consider $M \gtrsim 1$ TeV. This also evades bounds from single leptoquark production at HERA \cite{WICHMANN:2013voa}. Consequently, $|\lambda_{se}\lambda_{be}^*| \gtrsim 2 \cdot 10^{-3}$. Leptoquarks which couple to electrons can mediate t-channel di-jet production at an $e^+ e^-$ collider. Non-observation of any deviations at LEP requires 
\begin{align} |M/\lambda_{qe}| \gtrsim 10\, {\rm TeV}\ , \quad q=s,b
\label{eq:LEP}
\end{align} which is also easily satisfied.

Another bound can be derived from $B_s$ mixing. The interactions in \refeq{wunderbar} allow a box diagram with electrons and leptoquarks in the loop which gives rise to an operator of the form $\bar{b}s\bar{b}s$ with the complex coefficient 
\bea
\frac{\lambda_{se}\lambda_{be}^*}{16 \pi^2} \ \frac{\lambda_{se}\lambda_{be}^*}{M^2} \ .
\eea

Experimentally, the $B_s-\bar B_s$ mixing phase (defined relative to the SM phase in the amplitude for the gold-plated decay)
is bounded to be small, $0.00 \pm 0.07$ \cite{hfag}. Assuming a maximal CP phase in $\lambda_{se}\lambda_{be}^*$ this implies the bound
\begin{align}
|\lambda_{se}\lambda_{be}^* | 
\lesssim  0.07\, (24 {\rm TeV})^2\, \frac{(  V_{ts}^* V_{tb})^2 g^2 }{m_W^2} \sim 4 \ ,
\end{align}
where we also fixed $M^2/\lambda_{se}\lambda_{be}^*\simeq(24$ TeV$)^2$. It follows that $M \lesssim 48$ TeV and combining with
(\ref{eq:LEP}), $|\lambda_{qe}| \lesssim 5$. 
In the absence of CP violation, one still obtains a bound from the mass difference $|\Delta m_s^{\rm NP}/\Delta m_s^{\rm SM}| \lesssim 0.15 $ which is about a factor of two weaker because of  hadronic uncertainties \cite{Lenz:2011ti}.

We summarize the approximate boundaries of parameter space consistent with direct searches, $R_K$ and
$B_s$ mixing
\begin{align}
1\,  \mbox{TeV}  \lesssim M \lesssim 48 \,  \mbox{TeV} \, , \\
2 \cdot 10^{-3} \lesssim |\lambda_{se}\lambda_{be}^*  |  \lesssim 4 \, , \\
4 \cdot 10^{-4} \lesssim |\lambda_{qe}|  \lesssim 5 \, .
\end{align}
The last equation limits the hierarchy between the two couplings
$\lambda_{se}$ and $\lambda_{be}$.

There is also a constraint from the anomalous magnetic moment of the electron, which 
agrees with its SM prediction to a very high precision,  $\Delta a_e=-(10.5 \pm 8.1) \cdot 10^{-13}$ \cite{Giudice:2012ms}. Our model has a new one-loop contribution to the magnetic moment which is suppressed by the electron mass squared because of chiral symmetry
\begin{align}
\Delta a_e \sim \frac{|\lambda_{qe}|^2}{16 \pi^2} \ \frac{m_e^2}{M^2} \, , \quad q=s,b  \, .
\end{align}
This is much smaller than the present experimental uncertainty.

Predictions for other modes: 
$b \to s \nu \bar \nu$ processes can be mediated by the two operators
\begin{align} 
  {\cal{O}}_{L/R}^{\nu_\ell} & =  [\bar{s} \gamma_\mu P_{L/R} b] \, [ \bar{\nu}_\ell \gamma^\mu(1- \gamma_5) \nu_\ell]
\end{align}
in the low energy theory (\ref{eq:Heff}). In the SM
\bea
C_L^{\nu_\ell}|_{\rm SM} =-6.4\ , \quad C_R^{\nu_\ell}|_{\rm SM} \simeq 0\ .
\eea
In the leptoquark model the operator ${\cal{O}}_{RL}^e$ in \refeq{yukawabs} contains ${\cal{O}}_R^{\nu_e}$,
and we have 
\bea
C_R^{\nu_e} |_{\rm LQ} =C_9^{\prime e}= -C_{10}^{\prime e}
\simeq 0.5 \ , \quad
C_L^{\nu_e} |_{\rm LQ}  =0\ .
\eea

This predicts that the branching ratio of $\bar B \to \bar K \nu \nu $ which is
proportional to $\sum_{\nu_\ell} |C_L^{\nu_\ell} +C_R^{\nu_\ell}|^2$ 
is reduced by 5 percent relative to the SM one.
Note the sum over all three neutrino species but in our scenario only one, $\nu_e$, has BSM contributions.
On the other hand, the branching ratio of $\bar B \to \bar K^* \nu \nu$ would be enhanced relative to the SM
by about 5  percent because the dominant term in the decay rate is proportional to 
$|C_L^{\nu_\ell} -C_R^{\nu_\ell}|^2$ (with some uncertainty stemming from the relative size of form factors \cite{Buchalla:2000sk}).
The RL leptoquark contribution also enhances $F_L$, the fraction of longitudinally polarized $K^*$ in $\bar B \to \bar K^* \nu \nu$ relative to the SM by about two percent \cite{Altmannshofer:2009ma}.
The decays $\bar B \to \bar K \nu \nu$ and $\bar B \to \bar K^* \nu \nu$ have not been observed yet. Upper limits on their branching ratios are currently a factor of 3-4 ($K$) and 10 ($K^*$) above the SM predictions. These decays will be studied in the near future at the Belle II experiment at KEK. The inclusive mode $\bar B \to X_s \nu \nu$ is even more challenging experimentally. The enhancement of its branching ratio in this BSM scenario is below the permille level.

The $RL$ leptoquark model also induces contributions to the chirality flipped dipole operator ${\cal{O}}_7^\prime \propto m_b \bar s \sigma_{\mu \nu}F^{\mu \nu} P_L b $
through diagrams with the leptoquark and an electron running in a loop. This contributes to $b \to s\gamma$ and also to $b \to s \ell \ell$ decays
proportional to $\lambda_{se}\lambda_{be}^*/M^2$. This is the same combination of couplings and masses as in \refeq{yukawabs}, but suppressed by a loop factor relative to
$C_{RL}^e \sim 1$. The resulting fraction of ``wrong-sign" helicity photons (relative to the SM process) is then of the order few percent, in reach of  future high luminosity flavor factories with $75 {\rm ab}^{-1}$ \cite{Hitlin:2008gf}. More detailed study is needed to understand whether these ``wrong helicity" photon events from new physics can be separated from the respective ``wrong helicity" SM background, which arises at a similar level,  {\it i.e.} suppressed by $m_s/m_b$ relative to the dominant SM helicity at quark level. This is beyond the scope of our work.

\subsection{A model with a $LL$ operator for muons \label{sec:LLmu}}

We already showed in Section \ref{sec:opsA} that the single muonic operator $\mathcal{O}_{LL}^\mu$ can simultaneously explain both deviations in $R_K$ and $\bar B_s \to \mu \mu$. In fact, since the leptoquarks which we are considering are scalars, and since scalars (like the Higgs) might be expected to couple more strongly to the 2nd generation than to the 1st, it is natural to expect that the Wilson coefficients for muonic operators dominate over those for electrons. 

To construct a leptoquark model for $\mathcal{O}_{LL}^\mu$, note that it must involve both left-handed quarks and leptons.
Thus we write 
\bea
\label{eq:leptoLL}
\mathcal{L}=-\lambda_{b\mu}\, \phi^*\, q_3 \ell_2 -\lambda_{s\mu}\, \phi^*\, q_2 \ell_2 \ ,
\eea
where $q_i$ is the $i$-th generation left-handed quark doublet and $\ell_i$ is the $i$-th generation left-handed lepton doublet. These couplings require the leptoquark $\phi$ to have $(SU(3), SU(2))_{U(1)}$ quantum numbers $(3,1)_{-1/3}$ or $(3,3)_{-1/3}$, depending on how the $SU(2)$ indices in \refeq{leptoLL} are contracted. The $(3,1)_{-1/3}$ leptoquark couples down-type quarks only to neutrinos; it cannot generate the decays to muons that we are interested in. We therefore consider the $(3,3)_{-1/3}$ which mediates FCNCs with $|\Delta B|=|\Delta S|=1$ decays to muons as well as to neutrinos.

Integrating out the leptoquark and Fierz rearranging, we obtain flavor-preserving four-Fermi terms as well as
\bea
\label{eq:yukawaLLbs}
{\cal{H}}_{\rm eff}= 
- \frac{\lambda_{s\mu}^*\lambda_{b\mu}}{M^2} &&\!\!\!\!\!\!\left( \frac14[\bar{q_2} \tau^a \gamma^\mu P_{L} q_3] \,
[\bar{\ell_2}  \tau^a \gamma_\mu P_L \ell_2]  \right. \nonumber \\
&&\left.\!\!+ \frac34[\bar{q_2}  \gamma^\mu P_{L} q_3] \,
[\bar{\ell_2}  \gamma_\mu P_L \ell_2]  \right)
\eea
where $\tau^a$ are Pauli matrices contracted with the $SU(2)_L$ indices of the fermions. Since the fermions are $SU(2)_L$ doublets, these operators contain several different flavor-contractions for up- and down-type quarks, muons and muon neutrinos. In addition to the FCNC for $b \to s \mu \mu$ which was the goal of the model we also obtain 3 others
\bea
\label{eq:otherops}
[\bar s \gamma^\mu P_L b]\,[ \bar \mu \gamma_\mu P_L \mu]&,& \ \
\frac12[ \bar s \gamma^\mu P_L b] \, [\bar \nu_\mu \gamma_\mu P_L \nu_\mu] \, , \nonumber \\
\frac12[ \bar c \gamma^\mu P_L t] \, [ \bar \mu \gamma_\mu P_L \mu]&,& \ \
[ \bar c \gamma^\mu P_L t]\,[ \bar \nu_\mu \gamma_\mu P_L \nu_\mu] \, ,
\eea
all with the same coefficient. The two operators involving top quarks mediate top FCNC decays. Fixing the overall coefficient of the operator to explain the $R_K$ data, the top quark FCNC branching fraction is about $10^{-11}$, far too small to be observable. 

Moving on to the operator for b decays to muons, we obtain 
\bea
C_9^{\rm NP\mu}=-C_{10}^{\rm NP \mu}=\frac{\pi}{\alpha_{e}} \frac{\lambda_{s\mu}^*\lambda_{b\mu}}{V_{tb} V_{ts}^*}
\frac{\sqrt{2}}{2M^2G_F} \simeq -0.5 \ ,
\eea
where the last equality corresponds to the choice of Wilson coefficients which we determined as our benchmark point in Section \ref{sec:opsA}. Solving for the combination of free parameters in the model we find that we must choose $M^2 \simeq \lambda_{s\mu}^*\lambda_{b\mu} (48\, {\rm TeV})^2$.

Constraints on the parameter space of this model are very similar to the constraints of the electron model discussed in Section \ref{sec:RLe}. There is a bound from leptoquark pair production at the LHC, a bound from $B_s$ mixing,  and a bound from $g-2$ of the muon. These bounds are all easily satisfied for leptoquark masses between 1 and 48 TeV and $\sqrt{|\lambda_{s\mu}^*\lambda_{b\mu}|}\simeq M/(48\, {\rm TeV})$.

{} From \refeq{otherops} we see that the neutrino operator ${\cal{O}}_L^{\nu_\mu}$ is induced such that
\bea
C_R^{\nu_\mu} |_{\rm LQ} =0
 \ , \quad
C_L^{\nu_\mu} |_{\rm LQ}  =C_9^{\rm NP \mu}/2 \simeq -0.25\ .
\eea
This implies that the $\bar B \to \bar K^{(*)} \nu \nu$ and $\bar B \to X_s \nu \nu$ branching ratios are enhanced by 3\% 
whereas there is no effect on $F_L$.

In addition, there is a 1-loop induced contribution to the electromagnetic dipole operator ${\cal{O}}_7 \propto m_b \bar s \sigma_{\mu \nu}F^{\mu \nu} P_R b $. Given $C_{LL}^\mu \sim -1$, it implies an order few percent correction to the SM Wilson coefficient of ${\cal{O}}_7$. Besides in the global $|\Delta B|=|\Delta S|=1$ fits, this could be probed {\it e.g.} with the $b \to s \gamma$ branching ratio or the location of the zero of
$A_{\rm FB}(\bar B \to X_s \ell \ell)$. Future high luminosity flavor factories (with $75 {\rm ab}^{-1}$) are close to matching the requisite experimental precision  \cite{Hitlin:2008gf}.

\section{Summary \label{sec:con}}

Flavor physics can provide clues for physics at the weak scale and beyond. In this article we studied
 BSM physics that can affect the ratio $R_K$. A value of $R_K$ which differs from one would be a clean indication for  lepton-non-universal BSM physics which affects $b \to s ee$ and $b \to s \mu \mu$ transitions differently. Unlike  the individual $\bar B \to \bar K \ell \ell$, $\ell=e,\mu$  branching fractions, $R_K$ is
 essentially free of hadronic uncertainties, notably form factors.
 
 Anticipating that the current experimental situation holds up and a value of $R_K$ significantly smaller than one is confirmed, we explore possible new physics explanations. Interpretations with $bs \ell \ell$ tensor operators are already excluded by current data. Interpretations with (pseudo-)scalar operators are disfavored by data on $\bar B_s \to ee$, $\bar B_s \to \mu \mu$ and $\bar B \to \bar K ee$ decays. However, a fine-tuned possibility still survives which requires the  simultaneous presence of $O_{S,P}^{e}$ and the chirality-flipped $O_{S,P}^{e \prime}$. This scenario can be tested with an angular analysis of $\bar B \to \bar K ee$ decays.
 
 (Axial)-vector operators can provide an explanation of the $R_K$ measurement (\ref{eq:RKdata}). The effect could come from new physics coupling to muons, or electrons, or a combination thereof as in Eq.~(\ref{eq:Xlimit}) \cite{Das:2014sra}. In the near term, high statistics analyses of $\bar B \to \bar K^{(*)} \mu \mu$ and related decays at
 LHC(b) should clarify the situation in the muon channel.

We stress that the chiral nature of the SM fermion motivates the expectation that dimension six operators from new physics may be simplest in a chiral basis. We propose that global fits with chiral $SU(2)_L$-invariant
lepton currents be performed. It seems reasonable to assume that a single chiral operator dominates so
that the fit includes only a single parameter (the coefficient $C_{XY}$ of one of the ${\cal{O}}_{XY}$ with $X,Y \in L,R$). In the standard basis this would mean that one turns on only two of the 4 operators
${\cal {O}}_{9,10}^{(')}$ with one of the constraints
\begin{align}
C_9^{\rm NP \ell } = \pm C_{10}^{\rm NP \ell } \, , \quad C_9^{\rm NP \prime \ell } = \pm C_{10}^{\rm NP \prime \ell} \, .
\end{align}

We constructed two simple ``straw man" models with leptoquarks as examples for UV completions to the four-fermion operators with either muons or electrons.

We stress that the possibility of lepton-non-universal new physics strongly motivates related BSM searches: those with decays to ditau final states $b \to s \tau \tau$ and those with decays to $SU(2)_L$-partners,
$b \to s \nu \nu$. We give predictions for  di-neutrino modes, promising for the forthcoming Belle II
experiment, in Section
\ref{eq:particles}.

We further highlight Belle's preliminary results for the branching ratios of inclusive decays \cite{iijimaLP09},
\bea
{\cal{B}}(\bar B \to X_s e e)^{Belle}&=&(4.56 \pm 1.15^{+0.33}_{-0.40} )\cdot 10^{-6} \, , \nonumber  \\
{\cal{B}}(\bar B \to X_s \mu \mu)^{Belle}&=&(1.91 \pm 1.02^{+0.16}_{-0.18} )\cdot 10^{-6} 
\label{eq:Xsbelle}
\eea
with  $q^2 > 0.04 \mbox{ GeV}^2$. These branching ratios exhibit a similar enhancement of electrons versus muons as for $R_K$, although within sizable uncertainties: $R_{X}^{Belle} =0.42 \pm 0.25$. On the other hand, BaBar has obtained branching ratios in the same $q^2$-range (but with larger uncertainties \cite{hfag}) which are consistent between electrons and muons
\begin{align} \nonumber 
{\cal{B}}(\bar B \to X_s e e)^{Babar}=(6.0\pm 1.7 \pm 1.3 )\cdot 10^{-6} \, , \\
{\cal{B}}(\bar B \to X_s \mu \mu)^{Babar}=(5.0 \pm 2.8 \pm 1.2) \cdot 10^{-6} \, .
\label{eq:Xsbabar}
\end{align}
We used a combination of (\ref{eq:Xsbelle}) and (\ref{eq:Xsbabar}) as input for our analysis (\ref{eq:BXsll}). It would be desirable to obtain improved data on $R_{X}$, with lepton cuts corrected for \cite{Huber:2007vv}, to clarify this situation.

In addition, we emphasize that a high $q^2$-measurement of $R_K$ would be desirable to confirm or disprove lepton-non-universality in $b \to s$ decays.

Finally, if the leptoquarks are sufficiently light, then they can be produced in pairs at the LHC. It is natural to assume that the leptoquark couplings to to 3rd generation quarks might be larger than those to 2nd generation quarks, {\it  i.e.}~$\lambda_{b\ell} > \lambda_{s\ell}$. Then one expects decays to 1st and 2nd generation leptons with 3rd generation quarks. The two leptoquarks in the $SU(2)$ doublet of the RL model decay as 
\bea
\phi^{2/3} \to b\ e^+\ , \quad  \phi^{-1/3} \to b\ \nu
\eea
whereas the three leptoquarks in the $SU(2)$ triplet of the LL model decay as 
\bea
\phi^{2/3} &\to& t\ \nu \nonumber \\
\phi^{-1/3} &\to& b\ \nu \ , \  t\ \mu^- \nonumber \\
\phi^{-4/3} &\to& b\ \mu^- 
\eea
We emphasize that the resulting final states are currently not covered by most leptoquark searches because it is usually assumed (without theoretical basis) that leptoquark couplings involve only quarks and leptons of the same generation. But our $\phi$ scalars are neither 1st, 2nd, nor 3rd generation leptoquarks!

\vskip.2in
Note added: During the preparation of our manuscript preprint \cite{Alonso:2014csa} appeared which also 
points out relations imposed by $SU(2) \times U(1)$ invariance in the analysis of  $b \to s \ell \ell$ processes.

\acknowledgments
We are happy to thank Wolfgang Altmannshofer for useful discussions.
This work is supported in part by the Bundesministerium f\"ur Bildung und Forschung (GH) and by the US Department of Energy (MS).
We are grateful to the Aspen Center for Physics where this project was initiated for its hospitality and stimulating environment.


\end{document}